\input harvmac.tex
\noblackbox
\input epsf.tex
\overfullrule=0mm
\newcount\figno
\figno=0
\def\fig#1#2#3{
\par\begingroup\parindent=0pt\leftskip=1cm\rightskip=1cm\parindent=0pt
\baselineskip=11pt
\global\advance\figno by 1
\midinsert
\epsfxsize=#3
\centerline{\epsfbox{#2}}
{\bf Fig. \the\figno:} #1\par
\endinsert\endgroup\par
}
\def\figlabel#1{\xdef#1{\the\figno}}
\def\encadremath#1{\vbox{\hrule\hbox{\vrule\kern8pt\vbox{\kern8pt
\hbox{$\displaystyle #1$}\kern8pt}
\kern8pt\vrule}\hrule}}

\Title{\vbox{\baselineskip12pt
\hbox{USC-98-05}}}
{\vbox{\centerline{Multiparameter Integrable QFT's with $N$ bosons}}}

\centerline{ H.Saleur\foot{Packard Fellow}
 and  P.Simonetti}
\vskip2pt
\centerline{Department of Physics}
\centerline{University of Southern California}
\centerline{Los Angeles CA 90089-0484}
\vskip.3in
We introduce a new family of integrable theories with $N$ bosons and
$N$
freely adjustable mass parameters. These
theories restrict in particular limits to  the ``generalized
supersymmetric''
sine-Gordon models, as well as to  the flavor anisotropic chiral
Gross Neveu models (studied
recently by N. Andrei and collaborators). The scattering theory
involves
scalar particles that are no bound states, and bears an intriguing
resemblance
wih the results of a sharp cut-off analysis of the Thirring model
carried out by Korepin in (1980).  Various physical applications are
discussed. In particular,
we demonstrate that our theories are the appropriate continuum limit
of
integrable quantum spin chains with mixtures of spins.

\Date{4/98}

\newsec{Introduction}

A variety of low dimensional experimental condensed matter systems
have been
studied recently, that involve field theories with several bosons.
Examples
 include tunneling in quantum wires, where two bosons are necessary
to
describe the  charge and spin degrees of freedom of the electrons
\ref\KF{C. Kane, M. Fisher, Phys. Rev. B46 (1992) 15233.}, tunneling
between multiple edges in fractional quantum Hall devices
\ref\Andreas{C. Nayak, M. P. Fisher, A. W. W. Ludwig and H. H. Lin,
cond-mat/9710305.}, nanotubes and
two-leg ladders \ref\Nano{H. H. Lin, L. Balents and M. P. Fisher,
cond-mat/9801285.}, etc. Properties of interest in these systems are
usually non perturbative, and only a few techniques are available to
obtain quantitatively reliable results,
mostly conformal invariance and integrability.  The search for
integrable
 quantum field theories with several bosons is thus of some
importance.

The problem is, that besides the sine-Gordon model, most known
integrable
 bosonic theories are of little practical use: they are usually of
Toda type,
and involve real exponential of fields, that usually do not appear in
 a condensed matter context. Some exceptions to this unsatisfactory
situation are known: for instance, the double sine-Gordon model turns
out to
be exactly solvable for some values of the couplings \ref\VF{V. A.
Fateev,
 Nucl. Phys. B473 (1996) 509}, \ref\BL{A. P. Bukhvostov, L. N.
Lipatov, Nucl. Phys.
B180 (1981) 116.}, \ref\LSSII{F. Lesage, H. Saleur, P. Simonetti,
cond-mat/9707131.},
with potential applications to quantum wires.  Also,
the ``generalized supersymmetric'' extensions of the sine-Gordon
model
\ref\dBL{D. Bernard, A. Leclair, Comm. Math.
Phys. 142 (1991) 99.}  can be
rebosonized using standard bosonization formulas for the parafermions
\ref\GN{P. Griffin, D. Nemeschanksy, Nucl.Phys.B323 (1989) 545.}.
These
 theories  are
useful in the discussion of the multichannel Kondo model
\ref\Kondo{See for instance
A. C. Hewson, ``The Kondo problem
to heavy fermions'', Cambridge Studies in Magnetism, Cambridge (1993)
and references
therein; N. Andrei and C. Destri, Phys. Rev. Lett. 52 (1984) 364;
I. Affleck and A. W. W. Ludwig, Nucl. Phys. B360 (1991) 641.},
\ref\EK{V. Emery, S. Kivelson, Phys. Rev. B47 (1992) 10812.},
\ref\FG{M. Fabrizio,
 A. O. Gogolin, cond-mat/9407104.}; the ${\cal N}=1$
supersymmetric sine-Gordon model also appears in the context of
quantum wires \ref\LSSI{F. Lesage, H. Saleur, P. Simonetti,
cond-mat/9703220.}.

In this paper, we point out that there is a simple, integrable family
of
 theories extending the generalized supersymmetric sine-Gordon
models,
that involve $N$ bosons and have $N$ adjustable mass parameters. This
family
can be considered as an extension  of the flavor anisotropic Gross
Neveu models
that have been  studied in the last few years by N.Andrei and
collaborators(
mostly in the context of the channel anisotropic Kondo model
 \ref\NatanI{N. Andrei, A. Jerez, cond-mat/9412054}, \ref\NatanII{N.
Andrei,
 M. Douglas, A. Jerez, cond-mat/9502082}, \ref\NatanIII{N. Andrei, P.
Zinn-Justin,
 cond-mat/9801158.}), to the case where an anisotropy is introduced
both in
the color and flavor sectors. The case where the color anisotropy is
at the
special ``Toulouse'' value is of special interest for applications to
quantum
wires or dissipative brownian motion \ref\AOS{I. Affleck, M.
Oshikawa, H. Saleur,
in preparation}.

The models are presented in section 2, where integrability is proven
and various
 limiting cases discussed. The scattering theory is discussed in
section 3. The ``classical''
limit is analyzed in section 4, providing a general check of our
approach. In section 5,
the relation with quantum spin chains involving several species of
spins
is discussed. Some applications to impurity problems are discussed in
section 6.
Some final remarks are collected  in section 7.  The appendix
contains
numerous details on the numerical treatment both of the perturbation
theory
and of the TBA.

\newsec{Generalities}

\subsec{The integrable theories}

We consider a system of $N$ chiral bosons with propagators
\eqn\prop{\eqalign{<\phi_j(z)\phi_j(w)>&=-2{N-1\over N} \ln(z-w)\cr
<\phi_j(z)\phi_k(w)>&={2\over N} \ln(z-w),\ j\neq k,\cr}}
and introduce the following
fields:
\eqn\paraf{\Psi^{(j)}={1\over \sqrt{N}}\left(
\sum_{k=1}^{N} \omega^{jk} e^{i\phi_k}\right),}
where $\omega=e^{2i\pi/N}$. These fields  provide different
realizations
\GN\ of the fundamental parafermion \ref\FZ{V. A. Fateev, A.B.
Zamolodchikov, Sov.Phys.JETP 62 (1985) 215}
of $Z_N$ type. The  bosonic fields $\phi_j$  are not independent (one
can set
indeed $\phi_N=-\phi_1-\ldots-\phi_{N-1}$); they can be expressed in
terms of $N-1$ independent fields $\Phi_j$  obeying
\eqn\newprop{<\Phi_j(z)\Phi_k(w)>=-2\delta_{jk}\ln(z-w),}
by the transformation
\eqn\transf{\phi_j=e_j\bullet \Phi, j=1,\ldots, N,}
where the $\bullet$ denotes scalar product and the $e_j$ are weights
of the   fundamental
representation of  $SU(N)$\foot{That is, $e_1=\Lambda_1$,
$e_2=\Lambda_2-\Lambda_1$,
..., $e_{N-1}=\Lambda_{N-1}-\Lambda_{N-2}$, $e_N=-\Lambda_{N-1}$,
$\Lambda_i$ the fundamental weights of $SU(N)$ and $\Phi$ the $(N-1)$
dimensional vector of coordinates $\Phi_1,\ldots,\Phi_{N-1}$.}.

Introduce one additional  bosonic field, which has trivial
contractions with
the preceding ones, and obeys
\eqn\moreboson{<\Phi(z)\Phi(w)>=-{1\over 4\pi}\ln (z-w).}
Consider then the action (we assume all the $a_j$ are real positive
numbers)
\eqn\act{\eqalign{
A=&{1\over 2}\int dxdy\sum_{j=1}^{N-1} \left[\partial_\mu
\left(\Phi_j+\bar{\Phi}_j\right)\right]^2+\left[\partial_\mu\left(\Phi+\bar{\Phi}
\right)\right]^2\cr
&+\left(\sum_{j=1}^{N} a_j\Psi^{(j)}(z)\right)
\left(\sum_{j=1}^{N} a_j\bar{\Psi}^{(j)}(\bar{z})\right)
e^{i\beta\left[\Phi
(z)+\bar{\Phi}(\bar{z})\right]}+\hbox{ conjugate}.\cr}}
In the case where all the coefficients $a_j$ but one vanish, this is
the
action of the ``generalized supersymmetric'' sine-Gordon model, which
is known to be integrable \dBL.
We claim that this only a particular case of a more general
integrable
model, given by \act.

To establish this result, we first observe that the fields
$\Psi^{(j)}$ obey
the short distance expansions
\eqn\reli{\left[\Psi^{(j)}\right]^\dagger(z)\Psi^{(j)}(w)
\approx {1\over (z-w)^{{2(N-1)\over N}}}
\left[1+2{N+2\over N}(z-w)^2 T^{(j)}(w)+\ldots\right],}
where
$$
T^{(j)}(z)= {1\over N+2}\left(-{1\over 2} \sum_{k=1}^N
(\partial_z\phi_k)^2
+\sum_{k\neq l} \omega^{j(k-l)}e^{i(\phi_k-\phi_l)}\right),
$$
and
\eqn\relii{\left[\Psi^{(j)}\right]^\dagger(z)\Psi^{(k)}(w)\approx
{1\over (z-w)^{2{N-1\over N}}}\left[(z-w)J^{(jk)}(w)+\ldots\right],}
where
$$
J^{(jk)}=-{i\over N}\sum_{k=1}^N \omega^{-(k-j)l}\partial_z\phi_l
$$
We can then prove  integrability,
following
\ref\ABL{
C. Ahn, D. Bernard, A. Leclair, Nucl. Phys. B346 (1990) 409.}
, by establishing the existence of non local
conserved currents.  Introduce
\eqn\firstcons{{\cal J}^-(z)=\left(\sum_{j=1}^{N}
b_j\left[\Psi^{(j)}\right]^\dagger
(z)\right)
\exp\left(-i{4\pi\over \beta}{2\over N}\Phi(z)\right).}
then the short distance expansion of this
current with the first term in the action reads, for the chiral part,
$$
\eqalign{\left(\sum_{j=1}^{N} b_j\left[\Psi^{(j)}\right]^\dagger
(z)\right)\left(\sum_{k=1}^{N} a_k\Psi^{(j)}
(w)\right) \exp\left(-i{4\pi\over \beta}{2\over N}\Phi(z)\right)
\exp\left(i\beta\Phi(w)\right)\cr
\approx {1\over (z-w)^2}\left(\sum_{k=1}^N b_ka_k
+(z-w)\sum_{k\neq l} b_ka_l
J^{(kl)}(w)+\ldots\right)\exp\left(-{8i\pi\over \beta N}
\Phi(z)+i\beta\Phi(w)\right).\cr}
$$
The residue of the simple pole with thus be a total derivative iff
the factor of $(z-w)$ in the first bracket vanishes. This is
equivalent to
the condition
$$
\sum_{k\neq l} b_ka_l \left[\omega^{(k-l)m}-1\right]=0,
m=1,\ldots,N-1
$$
which always has solutions, since it is a system of $N-1$
equations with $N$ unknown \foot{The solution is easily expressed
using the matrix $N\times (N-1)$ matrix $M$ whose elements are
$M_{jk}=a_{j+k-1}$ by $b_j$ equal to the $j^{th}$ cofactor.}.

The short distance expansion of this current with the chiral part of
the second
term in the action has a leading term that goes as
$(z-w)^{-2/N}(z-w)^{2/N}$,
and thus no simple pole. Following the standard argument, the
expansion of ${\cal J}^-$ having a simple pole whose residue is a
total derivative, the non local charge $\int {\cal J}^-$ is conserved
to first order in the perturbation. For generic value of $\beta$, one
can then argue that this is true to any order in perturbation theory,
and, presumably, non perturbatively as well.

Another conserved current is easily found by complex conjugation:
\eqn\secondcons{{\cal J}^+(z)=\left(\sum_{j=1}^{N} b_j^*
\Psi^{(j)}
(z)\right)
\exp\left(i{4\pi\over \beta}{2\over N}\Phi(z)\right).}
The conservation of ${\cal J}^\pm$ then ensures integrability \ABL.

\subsec{The case $N=2$}

Let us discuss in more details the simplest example where $N=2$. In
that
case, the parafermions are self-conjugated (up to a sign). We set
\eqn\conv{\eqalign{\psi^{(1)}=-i\sqrt{2}\sin\phi_1= i\chi\cr
\psi^{(2)}=\sqrt{2}\cos\phi_1=\psi,\cr}}
where $\psi$ and $\chi$ are (real) Majorana fermions.  The
perturbative part of the action reads then
\eqn\pertact{(a\psi+ib\chi)(a\bar{\psi}+ib\bar{\chi})e^{i\beta\Phi}+
(a\psi-ib\chi)(a\bar{\psi}-ib\bar{\chi})e^{-i\beta\Phi},}
that is, regrouping terms
%
\eqn\pertacti{2\left[a^2\psi\bar{\psi}-b^2\chi\bar{\chi}\right]\cos\beta\Phi+
2ab\left(\psi\bar{\chi}+\chi\bar{\psi}\right)\sin\beta\Phi.}
The non local conserved currents read then
\eqn\nonloci{\eqalign{{\cal J}^+(z)=&(a\psi-ib\chi)
\exp\left(i{4\pi\over \beta}\Phi\right)\cr
{\cal J}^-(z)=&(a\psi+ib\chi)
\exp\left(-i{4\pi\over \beta}\Phi\right).\cr}}

If $a=0$ or $b=0$, the action reduces to the one of the
supersymmetric
sine-Gordon model (with an additional, decoupled, Majorana fermion).
If $a= b$,
the combinations appearing in the action become  Dirac fermions,
$\psi+i\chi=\sqrt{2}
e^{i\phi_1}$. We can thus reexponentiate them, to write the
perturbing term as $a \cos(\beta\Phi+\phi_1)$, so the model is
equivalent
to a sine-Gordon model, at a coupling constant $\beta'$ with
${(\beta')^2\over 8\pi}={\beta^2\over 8\pi}+{1\over 2}$.

The currents ${\cal J}^\pm$ have a fractional
spin $s={1\over \gamma}$,  where
\eqn\gamdef{\gamma= {2\beta^2\over 4\pi -\beta^2}.}
In the case $a=0$ or $b=0$, the currents are generators
of the algebra  $\widehat{sl(2)_q}$ \ABL,  with deformation parameter
$q=-e^{-i\pi/\gamma}$. In the general case however, they  do not form
a closed algebra.

To understand the situation a little better, it is useful to
go to the $SU(2)$ symmetric point $\beta^2=4\pi$. There are
two underlying level one algebras, with generators
\eqn\firstalg{\eqalign{J_1^+=&{\psi+i\chi\over
\sqrt{2}}e^{i\sqrt{4\pi}\Phi}\cr
J_1^-=&{\psi-i\chi\over \sqrt{2}}e^{-i\sqrt{4\pi}\Phi}\cr
J_1^3=&i\left(-\psi\chi+\sqrt{4\pi}\partial\Phi\right)\cr}}
and
\eqn\secondalg{\eqalign{J_2^+=&{\psi-i\chi\over
\sqrt{2}}e^{i\sqrt{4\pi}\Phi}\cr
J_2^-=&{\psi+i\chi\over \sqrt{2}}e^{-i\sqrt{4\pi}\Phi}\cr
J_2^3=&i\left(\psi\chi+\sqrt{4\pi}\partial\Phi\right),\cr}}
and all short distance expansions between operators of different
algebras
are non singular. The sum $J_1+J_2$ provides a level two
representation. For general $a,b$,
the currents can be written as combinations of the $J_1$ and $J_2$.
Setting $a=\mu+\lambda$ and $b=\mu-\lambda$, we have
$$
\eqalign{{\cal J}^+=&\lambda J_1^++\mu J_2^+\cr
{\cal J}^-=&\lambda J_1^-+\mu J_2^-.\cr}
$$
This is suggestive of a system where two flavors of fermionic
currents
are combined in a flavor anisotropic fashion. Indeed, introduce
new bosons defined by
\eqn\moretransf{\eqalign{-\phi_1+\sqrt{4\pi}\Phi=&\varphi_1\cr
\phi_1+\sqrt{4\pi}\Phi=&\varphi_2,\cr}}
the perturbing term is then proportional to
$$
\lambda^2\cos(\varphi_1+\bar{\varphi}_1)+\mu^2\cos(\varphi_2+\bar{\varphi}_2)
+\lambda\mu\cos(\varphi_1+\bar{\varphi_2})+\lambda\mu\cos(\bar{\varphi}_1+
\varphi_2)
$$
This is the abelian bosonized form of a Gross Neveu type interaction
with flavor anisotropy
$$
\left(\lambda J_1^x+\mu J_2^x\right)\times
\left(\lambda \bar{J}_1^x+\mu \bar{J}_2^x\right) + (x\to y)
$$
The $zz$ term is missing in this interaction - it is well known that
this term is generated under renormalization \BL. Away from the
$SU(2)$ point,
one can similarly consider our model as a color and flavor
anisotropic chiral Gross Neveu model (upon bosonization, this model
gives rise to $4$ independent fields, but only $2$ of them appear in
the interaction due to chirality, the other $2$ ones contributing
free parts to the action).

\subsec{Arbitrary $N$}

The previous discussison easily extends to other values of $N$.
The currents ${\cal J}^\pm$ have a fractional
spin $s={1\over \gamma}$,  where
\eqn\gamdef{\gamma= {N\beta^2/ 8\pi\over {1\over N} -{\beta^2\over
8\pi}}.}
In the case $a=0$ or $b=0$, the currents are generators
of the algebra  $\widehat{sl(2)_q}$  with deformation parameter
$q=-e^{-i\pi/\gamma}$. In the general case, they  do not form a
closed algebra.
In the limit $\beta^2=4\pi$, they can be expressed as combinations
of $N$ generators belonging to $N$ different realizations of a level
1
$SU(2)$ algebra. The rebosonized action is a Gross Neveu models
with two colors and $N$ flavors, and flavor anisotropy, with an
interaction
term of the form
$$
\left(\sum_{j=1}^N\lambda_j J_j^x\right)\times
\left(\sum_{j=1}^N\lambda_j \bar{J}_j^x\right) + (x\to y)
$$
where the anisotropy coefficients $\lambda_j$ are related with the
terms in the original action by
\eqn\lamdef{
\lambda_j=\sum_{k=1}^N \omega^{jk} a_k.}
That the Gross Neveu model with flavor anisotropy is integrable has
been
pointed out several years ago in fact, in  \NatanI,\NatanII.
Integrability is
established there by direct diagonalization of the bare hamiltonian
together
with ``dynamical fusion''. An intriguing feature is that the proof
presented in
\NatanII, strictly speaking, works only for the case $N=2$ (and some
subcases
of special flavor anisotropy for larger $N$). The reason is, that in
the approach
 of
 \NatanII, the bare particles must have a bare flavor scattering
matrix
that is a solution of the  Yang Baxter equation. In the flavor
isotropic case,
this $S$ matrix
is the standard $SU(N)$ R-matrix; but, as far as we know, there is no
way
to deform this R matrix in a non trivial way by introducing $N-1$
independent
anisotropic parameters - all available single and multiparameters
quantum
group approaches still explore a very small subset of all the
possible flavor
anisotropies. On the other hand, from the point of view we have
adopted (that deals
directly with the renormalized action), all flavor anisotropies play
equivalent
roles, and integrability appears generally true. The argument also
extends
straightforwardly to the case of color anisotropy, not considered in
\NatanI.

Of course, a particular choice of anisotropy is when one of the
coefficients
$\lambda_j$ vanishes exactly,  in which case
the modele reduces to one with $N-1$ flavors. In the case of general
$\beta$,
the same conditon, say $\lambda_{N}=0$ leads, using formula \paraf,
to
a problem where the field $\phi_N$ has disappeared from the action.
We are then left with a set of $N-1$ fields satisfying
\eqn\propi{\eqalign{<\phi_j(z)\phi_j(w)>&=-2{N-1\over N} \ln(z-w)\cr
&=
-2{N-2\over N-1}\ln(z-w)-{2\over N(N-1)}\ln(z-w)\cr
<\phi_j(z)\phi_k(w)>&={2\over N} \ln(z-w)\cr&=
{2\over N-1}\ln(z-w)-{2\over N(N-1)}\ln(z-w),j\neq k\cr}}
We can then write $\phi_j=\phi'_j+\Phi'$ where there are $N-1$ fields
$\phi'_j$ satisfying relations similar to \propi\ but with the
replacement
$N\to N-1$. The problem is then equivalent to the case $N\to N-1$,
but with a shift of the leftover exponential, ${\beta^2\over 8\pi}\to
{\beta^2\over 8\pi}+{1\over N(N-1)}$.

\newsec{Conjectured scattering theory}

\subsec{The thermodynamic Bethe ansatz}

Though there are arguments based on symmetry to infer what the
scattering theory
should look like, the approach we  use is to first conjecture a set
of thermodynamic Bethe ansatz equations (TBA) to compute the free
energy of the
1+1 quantum field theory associated with the action \act\ at
temperature $T$.  We parametrize
\eqn\param{{\beta^2\over 8\pi}= {\gamma\over N(N+\gamma)},}
and consider  the case  $\gamma$ an integer. Our conjecture is as
follows.
Introduce
the TBA diagram
\bigskip
\noindent
\centerline{\hbox{\rlap
{\raise28pt\hbox{$\hskip6.7cm\bigcirc\hskip.25cm
\gamma+N-1$}}
\rlap{\lower27pt\hbox{$\hskip6.6cm\bigcirc\hskip.3cm \gamma+N$}}
\rlap{\raise15pt\hbox{$\hskip6.3cm\Big/$}}
\rlap{\lower14pt\hbox{$\hskip6.2cm\Big\backslash$}}
\rlap{\raise15pt\hbox{$\hskip.2cm 1\hskip.9cm 2\hskip1.3cm N
\hskip.5cm N+1
$}}
$\bigotimes$------$\bigotimes$-- -- --
--$\bigotimes$------$\bigcirc$-- --
--$\bigcirc$------$\bigcirc$\hskip.3cm $\gamma+N-2$
}}
\bigskip
\bigskip
\noindent
with incidence matrix $N_{jk}$ such that $N_{jk}=1$ if the nodes $j$
and $k$ are connected, and $0$
otherwise (in particular $N_{jj}=0$). With this diagram, we
associate the set of
pseudo energies
(one for each node) solution of the system ($R=1/T$, $T$ the
temperature)
\eqn\tba{\epsilon_j=\sum_{k=1}^N\delta_{jk}m_kR\cosh\theta-\sum_{k}
N_{jk}\int
{d\theta'\over 2\pi}{1\over \cosh{(\theta-\theta')}}
\ln\left(1+e^{-\epsilon_k(\theta')}\right).}
The free energy reads then
\eqn\freeen{F=-{T\over 2\pi}\sum_{k=1}^N\int {d\theta\over 2\pi}
m_k\cosh\theta \ln\left(1+e^{-\epsilon_k(\theta)}\right).}

In the foregoing equations, the $m_k$ are a set of masses which
depend on the couplings
$a_k$ in
the bare action. By dimensional analysis,
$[a_k]=[\hbox{length}]^{{\beta^2\over
4\pi}-{2\over N}}$. Therefore, we have
\eqn\mystery{m_k=G_k(a_1,\ldots, a_N)=a_1^{ N+\gamma
\over 2}F_k(a_2/a_1,\ldots, a_N/a_1).}
The $G_k$ are  homogeneous functions of the couplings $a_k$. Some
properties of these functions are known before hand of course. They
are
symmetric functions of their arguments. If all the $a_k$ but one
vanish,
we know that the problem becomes equivalent to the $N^{th}$
supersymmetric
sine-Gordon model, and therefore, from known results \ABL,
\ref\Zamo{Al. B.
Zamolodchikov, Nucl. Phys. B385 (1991) 497.}, all the masses but the
$N^{th}$ one must vanish.
This
means that $G_j=0$, $j=1, \ldots, N-1$, when all the $a_k$ but one
vanish.
Also, we know that the $N^{th}$ mass vanishes when one of the fields
decouples, ie when one of the coefficients \lamdef\ vanishes. More
generally,
the  masses $m_k\ldots m_N$  vanish when $N-k+1$ of these
coefficients vanish. We will get back to the determination of the
functions $G_k$ below.

The evidence for the TBA comes first from the compatibility with all
 limiting cases. Moreover, the analysis of the $Y$ system
\Zamo\ associated with it shows that the
dimension of the UV perturbing operator is always the same as
in the generalized supersymmetric case (it does not depend on the
number of massive nodes), ie $h={\gamma+N-1\over
\gamma+N}={\beta^2\over 8\pi}+{N-1\over N}$ as desired. Finally, the
central charge, in the generic case when all the $m_k$ are non zero
is simply equal to the number of massive nodes, ie $c=N$. This is
easily checked. Using standard formulas, the central charge is
expressed in terms of the solutions of the system \tba\ as $T\to 0$
and $T\to \infty$. In the first case, the $N$ first $\epsilon$'s are
all infinite, the others follow from
$$
\eqalign{x_j&=e^{-\epsilon_j}=(j+1)^2-1, j=N+1,\ldots, N+\gamma-2\cr
x_{N+\gamma-1}&=x_{N+\gamma}=\gamma-1.}
$$
In the second case, one has to solve the same system with more nodes,
ie
$$
\eqalign{y_j&=e^{-\epsilon_j}=(j+1)^2-1, j=1,\ldots, N+\gamma-2\cr
y_{N+\gamma-1}&=y_{N+\gamma}=N+\gamma-1.}
$$
The central charge is then (here $L$ designates the Euler
dilogarithm) \Zamo\
\eqn\centrch{c={6\over \pi^2}\sum L\left({y\over
1+y}\right)-L\left({x\over 1+x}\right).}
For a $D$ diagram, as $T\to\infty$,  the sum ${6\over \pi^2}\sum
L\left({y\over 1+y}\right)$  is equal to the number of nodes minus
one, so the central charge is simply, from $L(1)={\pi^2\over 6}$,
equal to the number of massive nodes, ie $c=N$ indeed.

\subsec{Scattering theory}

The scattering theory associated with this TBA is very simple. One
introduces a
set of $N-1$ scalar massive particles with masses $m_1,\ldots,
m_{N-1}$. One also introduces a pair  soliton/antisoliton with masses
$m_N$. The latter
scatter with the usual sine-Gordon S matrix that corresponds to the
quantum group parameter introduced above, $q=-e^{-i\pi/\gamma}$ - it
is the same as
the S matrix of an ordinary sine-Gordon model at coupling
${\beta_{eq}^2\over 8\pi}={\gamma\over \gamma+1}$. The scalar
particle of label $k$ scatters trivially with all particles, except
the ones of label $k\pm 1$, with which it
scatters with the CDD factor $S=i\tanh\left({\theta\over
2}-{i\pi\over 4}\right)$. When $k=N-1$, the particle scatters with
the soliton and antisolitons with the same CDD factor. It is
important to
stress that the sine-Gordon S-matrix considered here has no poles in
the physical
strip: the scalar particles are {\sl not} bound states of the soliton
and
 antisoliton.

Remarkably, a scattering theory built with
similar ingredients appears in a paper by Korepin \ref\K{V. Korepin,
Comm. Math. Phys. 76 (1980) 165.}. There, the author discusses
the  Thirring model in the repulsive
regime, using a sharp cut-off regularization. For a coupling
corresponding to a sine-Gordon parameter ${\beta^2\over 8\pi}
\in \left[{l\over l+1},{l+1\over l+2}\right]$, he finds, in addition
to
the soliton and antisoliton, a spectrum made of $(l-1)$
neutral particles, with same S-matrices as ours, but where  all the
masses
are uniquely
determined as a function of $l$ (in particular, the soliton mass
becomes infinite
when ${\beta^2\over 8\pi}={l+1\over l+2}$). The soliton and
antisoliton
scatter through a sine-gordon S matrix
with a renormalized $\left({\beta^2\over
8\pi}\right)_l={1-l(1-\beta^2/8\pi)\over
1-(l-1)(1-\beta^2/8\pi)}$ parameter.

The relation with our problem,
if any, is not clear. It is usually admitted that the in the
repulsive regime of SG, the quantum theory must be defined with care,
and depends on the cut-off. For
smoother cut-offs, as well as XXZ type regularizations, the results
of \K\
are not supposed to hold, and the standard description of
\ref\ZZ{A. B. Zamolodchikov and Al. B.
Zamolodchikov, Annals of Physics 120 (1979) 253.}
 with only the soliton and antisoliton to be correct.

Observe that the sine-Gordon part
of the
S matrix commutes with the quantum algebra $\hat{sl(2)}_q$. While the
non
local conserved charges ${\cal Q}^\pm$ have commutation relations
that do not close,
they can be expresed as combinations of $N$ basic charges generating
$\hat{sl(2)}_q$. A simple representation of the algebra generated by
the
${\cal Q}^\pm$ is then obtained by identifying all these charges -
it is realized on the
multiparticle soliton antisoliton states
as in the usual sine-Gordon model \ABL. In the general case, there
are no other conserved charges. If $N=2$ for instance,
another, local, conserved charge appears when eg $b=0$, since
the local currents ${\cal G}=\psi\partial\Phi$ is conserved \ABL. But
away from
this value (and $a=0$ of course), this is not true. We must thus
complete the S matrix by a sector with no apparent symmetries, that
does
not spoil the Yang Baxter equation: besides ``vertex'' and ``RSOS''
type
solutions, the only available choice is a set of particles with
diagonal scattering. Requiring the central charge to be $N$,
and the TBA to restrict to the known ones when some of the
couplings vanish (and additional symmetries appear) seems to leave
no choice but our result.

Still another check comes from discussing the limit $\beta\to 0$, to
which we turn now.

\newsec{The ``classical'' limit.}

We call here classical the limit where the sine-Gordon part of the
scattering theory
becomes identical with the classical one, that is $\beta\to
0,\gamma\to 0$.

\subsec{The TBA at the reflectionless points}

So far we wrote a TBA in the simplest case $\gamma$ an integer, for
which
${\beta^2\over 8\pi}\geq{1\over N(N+1)}$, corresponding
 to the SG
component of the S-matrix being in the repulsive regime. To approach
$\beta=0$,
we consider instead the cases
$\gamma={1\over n}$, that is ${\beta^2\over 8\pi}={1\over N(nN+1)}$:
the SG S-matrix is now in
the attractive regime,
at the so called reflectionless points. This means that the spectrum
has
to be completed by the bound states of solitons and antisolitons,
 the $n-1$ breathers of masses $2m\sin\left(j{\pi\over 2n}\right)$,
where $m$
is the soliton mass.  We denote  $2 m_1
\sin\left({\pi\over 2n}\right),\ldots, 2 m_{N-1} \sin\left({\pi\over
2n}\right)$ the masses of the scalar particles (recall these are not
bound states;
the factor $2\sin\left({\pi\over 2n}\right)$ in their masses is
introduced
for convenience only). By building the complete scattering
theory using bootstrap and fusion, one finds the following TBA
equations.

 One first
has equations for the right hand side of the diagram, that look like
the standard ones for the attractive regime of sine-Gordon
\eqn\rightside{\epsilon_j=\sum_k N_{jk}
K*\ln\left(1+e^{\epsilon_k}\right),  N+1\leq j\leq N+n.}
There is then a central part involving the nodes $N-1$ and $N$, with
\eqn\middi{\epsilon_{N}=K*\left[\ln\left(1+e^{\epsilon_{N+1}}\right)-
\ln\left(1+e^{-\epsilon_{N-1}}\right)\right],}
and
%
\eqn\middii{\eqalign{\epsilon_{N-1}=&\left[2m_{N-1}\cos\left({\pi\over
2n}\right)+ m\right] \tan\left({\pi\over 2n}\right){\cosh\theta\over
T}\cr
&-K*\ln\left(1+e^{\epsilon_N}\right)-K''
*\left(1+e^{-\epsilon_{N-1}}\right)-K'*\left(1+e^{-\epsilon_{N-2}}\right).\cr}}
Finally, the left hand side of the diagram looks in turn like the
usual
repulsive  one
\eqn\midiii{\epsilon_j=2m_j\sin\left({\pi\over 2n}\right)
{\cosh\theta\over T}-\sum_k N_{jk}
K'*\left(1+e^{-\epsilon_k}\right), j\leq N-2.}
These equations can be conveniently encoded in the diagram

\bigskip
\noindent
\centerline{\hbox{
\rlap{\lower8.5pt\hbox{$\hskip4.28cm\!\!\!\ast\!\!\!\bigcup\!\!\!\ast$}}
\rlap{\raise27pt\hbox{$\hskip7.75cm\bigotimes\hskip.25cmN+n-1$}}
\rlap{\lower27pt\hbox{$\hskip7.55cm\bigotimes\hskip.3cm N+n $}}
\rlap{\raise13pt\hbox{$\hskip7.4cm\Big/$}}
\rlap{\lower14pt\hbox{$\hskip7.25cm\Big\backslash$}}
\rlap{\raise13pt\hbox{$\hskip.3cm1\hskip.9cm 2\hskip1.7cm
N-1\hskip.5cm N$}}
 \hskip.08cm $\bigotimes$---$\!\!/\!\!$---$\bigotimes$-- -- --
$\bigotimes$---$\!\!/\!\!$---$\bigotimes$------$\bigotimes$-- -- --
$\bigotimes$------$\bigotimes$\hskip.3cm $N+n-2$
}}
\bigskip
\bigskip
\noindent

%
%

The asymptotic conditions for the attractive part are
\eqn\asymptmore{\eqalign{\epsilon_j&\approx
2m\sin\left[{(j-N+1)\pi\over 2n}\right]{\cosh\theta\over T}, N\leq
j\leq n+N-2\cr
\epsilon_{n+N-1}&\approx \epsilon_{n+N}\approx m{\cosh\theta\over
T}.\cr}}
Introducing the fourier transform $g(\theta)=\int_{-\infty}^\infty
{d\omega\over 2\pi} e^{-i\omega\theta}\tilde{g}(\omega)$, one has
$\tilde{K}={1\over 2\cosh(\pi\omega/2n)}$ (the standard
attractive kernel),   $\tilde{K}'={1\over
2\cosh(\pi\omega/2)}$ (the
standard repulsive kernel),
$\tilde{K}''= {\cosh(n-1)\pi\omega/2n\over
2\cosh\pi\omega/2n\cosh\pi\omega/2}$.

For $T\to 0$, we have $x_j=0$, since now all nodes are massive.  For
$T\to\infty$, we have
\eqn\sols{\eqalign{e^{-\epsilon_j}=&(j+1)^2-1, j\leq N-1\cr
e^{\epsilon_j}=&\left(j-N+1+{1\over N}\right)^2-1,N\leq j\leq
N+n-2\cr
e^{\epsilon_{N+n-1}}=&e^{\epsilon_{N+n}}=n+{1\over N}-1.\cr}}
The central charge is thus
\eqn\cc{
c={6\over \pi^2}
\left\{
2L\left({1\over n+{1\over N}}\right)+\sum_{j=1}^{n-1}
L\left[{1\over \left(j+{1\over N}\right)^2}\right]
+\sum_{j=1}^{N-1} L\left[1-{1\over (j+1)^2}\right]
\right\}=N.}

\subsec{The classical limit}

The ``classical limit'' is obtained  by letting
$\beta\to 0$. In our TBA, this means $n\to \infty$. To get non
trivial results then, we scale the soliton mass $m$ with $n$, so the
mass of the first breather
remains finite. Similarly, we assume that the parameters $m_j$ are
also scaled with $n$, so that the masses of the
scalar particles remain finite.  Our computation
follows the general strategy of \ref\Ts{A. M. Tsvelick and P. B.
Wiegmann,
Adv. in Physics 32 (1983) 331.}, \ref\Fow{
M. Fowler, Phys. Rev. B26 (1982) 2514.}. In that limit,
it is convenient to introduce the new notation $\kappa_j=
\epsilon_{j+N-1}$, $j\geq 1$. When $n\to\infty$, the kernels  $K$ and
$K''$ become delta functions, (we set $K'=s$) and  one finds
the general solution
$$
e^{\kappa_j}+1=\left({aA^j-a^{-1}A^{-j}\over A-A^{-1}}\right)^2
$$
The constant $A$ follows from the knowledge of mass terms,
$A=e^{C/2T}, C={m\pi\over n }\cosh\theta$, while the constant $a$
depends on $\epsilon_{N-1}$:
$$
1+e^{-\epsilon_{N-1}}=\left({A-A^{-1}\over a-a^{-1}}\right)^2
$$
In that limit, the equation satisfied by $\epsilon_{N-1}$ is
$$
\epsilon_{N-1}=-s*\ln\left(1+e^{-\epsilon_{N-2}}\right)
-{1\over 2}\ln\left(1+e^{\kappa_1}\right)-{1\over 2}
\ln\left(1+e^{-\epsilon_{N-1}}\right)+\left(\Lambda+{1\over
2}\right){C\over T}
$$
where $\Lambda={m_{N-1}\over m}$.

Let us then define
$$
\epsilon'_{N-1}=s*\ln\left(1+e^{\epsilon_{N-2}}\right)+\Lambda
{C\over T}.
$$
By simple algebra, one finds that the following holds
$$
1+e^{-\epsilon_{N-1}}=\left(1+e^{-\epsilon_N}\right)\left(1+e^{-\epsilon'_{N-1}}\right)
$$
together with
$$
\epsilon_N=(\Lambda+1){C\over
T}-s*\ln\left(1+e^{-\epsilon_{N-2}}\right)
$$

We can thus trade completely the right hand side of the
diagram for an additional node, and get the equations (recall
$s=K'$,$\tilde{s}=
{1\over 2\cosh(\pi\omega/2)}$)
\eqn\class{\epsilon_{j}={M_j\cosh\theta\over T}-\sum_k N_{jk}
s*\ln\left(1+e^{-\epsilon_k}\right), j=1,\ldots,N,}
where the TBA diagram is a D diagram with $N$ nodes

\bigskip
\noindent
\centerline{\hbox{\rlap{\raise28pt\hbox{$\hskip5.6cm\bigotimes\hskip.25cm N-1$}}
\rlap{\lower27pt\hbox{$\hskip5.5cm\bigotimes\hskip.3cm N$}}
\rlap{\raise15pt\hbox{$\hskip5.3cm\Big/$}}
\rlap{\lower14pt\hbox{$\hskip5.2cm\Big\backslash$}}
\rlap{\raise15pt\hbox{$1\hskip1cm 2\hskip1.5cm \hskip.8cm N-3$}}
$\bigotimes$------$\bigotimes$-- -- --
--$\bigotimes$-- -- --$\bigotimes$------$\bigotimes$\hskip.5cm $N-2$
}}

\bigskip
\noindent and  masses,
in that limit are  $M_1= m_1{\pi\over n},\ldots,
M_{N-1}=m_{N-1}{\pi\over n}$,
and $M_N=M_{N-1}+m{\pi\over n}$.

We can now compute the free energy. Its general expression is
\eqn\freen{{F\over T}=-\sum_{j=1}^\infty j\int {d\theta\over 2\pi}
C(\theta)\ln\left(1+e^{-\kappa_j}\right)-\sum_{j=1}^{N-1}
{m_j\over m}\int {d\theta\over 2\pi} C(\theta)
\ln\left(1+e^{-\epsilon_j}\right).}
By using the basic TBA equation
$$
\kappa_j\approx {jC\over T}+2\sum_{k=1}^\infty
k\ln\left(1+e^{-\kappa_k}\right)
-\ln\left(1+e^{-\epsilon_{N-1}}\right),
$$
one finds after a few simple manipulations
\eqn\finalfree{{F\over T}=-\sum_{j=1}^N \int {d\theta\over 2\pi} M_j
\cosh\theta \ln\left(1+e^{-\epsilon_j}\right)+\int {d\theta\over
2\pi} m\cosh(\theta)
\ln\left(1-e^{-M\cosh\theta/T}\right).}
Here, the new mass $M_N=M_{N-1}+{m\pi\over n}$ as before, and
$M={m\pi\over n}$.

The results of the classical limit are therefore equations \class\
and \finalfree.
This means, the classical limit is made up of
 $N$
particles scattering in a non trivial way, plus a decoupled free
boson.

In the particular case
of the generalized supersymmetric sine-Gordon model, all the masses
$M_j$ but the $N^{th}$ one vanish: the system
 \class\ reproduces the well known TBA
for $Z_N$ field theories perturbed by the parafermion field
\ref\Fat{V. A. Fateev, Int.J.Mod.Phys. A6 (1991) 2109.}, as is
expected
from letting $\beta\to 0$ in the action.

In the particular case $N=2$, there is no node $N-2$, and the TBA
system
is trivial. The system decouples into two free fermions of masses
${m_1\pi\over n}$ and $(m_1+m){\pi\over n}$,  and a free boson of
mass ${m\pi\over n}$. This is again expected from the action, and  a
rather
 non trivial check from the
point of view of the TBA. The mass for the  free boson certainly
arises from counter terms analogous to ones
arising in  the ${\cal N}=1$ supersymmmetric action \ref\SM{S.
Sengupta, P. Majumdar, Phys. Rev. B33 (1986) 3138.} (for more
discussion
about this, see next section).
It also follows that, in the classical limit, the correspondence
between
masses in the TBA and bare couplings goes, assuming $a^2\leq b^2$, as

\eqn\classicmasses{\eqalign{m_1\propto& a^2\cr
m\propto &(b^2-a^2),\cr}}

where $m$ is the mass of the first breather, and $m_1$ the mass of
the scalar particle. In the more general case, the decoupled  free
boson presumably gets
its mass
from a counter term still analogous to what happens in the
supersymmetric
case \SM. The rest of the TBA corresponds to a non trivial theory,
with
$N$ species of $Z_N$ parafermions interacting. It is interesting
to discuss in more details the case $N=3$. There, the TBA is based on
the diagram

\bigskip
\noindent
\centerline{\hbox{\rlap{\raise15pt\hbox{$\hskip .2cm 1\hskip.9cm
2\hskip.9cm 3$}}
$\bigotimes$------$\bigotimes$------$\bigotimes$ }}

\bigskip
\noindent with central charge $c=2$. Observe that this value can be
obtained
by $2={1\over 2}+{7\over 10}+{4\over 5}$, corresponding to the sum of
the central
charges for the Ising, tricritical and tetracritical Ising model (or,
alternatively, the Ising and tricritical Ising models, and the 3
state Potts
model). Remarkably, the weight $h={2\over 3}$ of our three
parafermions $\Psi^{(j)}$ can be recovered by using
fields of these minimal conformal field theories. The three fields
$\Phi_{13}^{(4/5)}$, $\Phi_{13}^{(7/10)}\Phi_{33}^{(4/5)}$
and $\Phi_{13}^{(1/2)}\Phi_{33}^{(7/10)}\Phi_{33}^{(4/5)}$ (where
lower labels are labels of the Kac table, upper labels
are the central chagres) do have
conformal weights $h={2\over 3}$. One might be tempted to infer
that the foregoing TBA also describes the perturbation
of this product of three theories by a combination
of these three fields. This cannot be true however: it is easy to
check that the operator algebra of these three fields cannot
be reproduced by using our parafermion fields only, due for instance
to the  appearance of powers $1/15$ and $19/15$. More generally,
a TBA like

\bigskip
\noindent
\centerline{\hbox{\rlap{\raise15pt\hbox{$\hskip .2cm 1\hskip.9cm
2\hskip 1.cm M$}}
$\bigotimes$------$\bigotimes$-- -- --$\bigotimes$ }}

\bigskip
\noindent has a central charge that can be written as
$c=c_1+c_2+\ldots c_M$,
where $c_M=1-{6\over (M+2)(M+3)}$. The conformal weight
of the perturbing operator is $h={M-1\over M}$, which
can be reproduced by $M$ fields of the form
$\Phi_{13}^{(c_m)}\Phi_{33}^{(c_{m+1})}\ldots\Phi_{33}^{c_M}$,
$m=1,\ldots,M$.
While
it
is tempting to speculate that ther TBA does describe the product of
$M$
minimal models perturbed by a combination of these fields,
this result does not seem to be  true.

\newsec{Numerical check}

To check the validity of the TBA besides the qualitative features we
just
 discussed, one needs to compare the result for the free energy
\freeen\ to
   perturbative computations. We restrict here to the simplest case
$N=2$.
Because the action \pertacti\ has two free parameters, reflected in
the existence
 of the two masses $m_1$ and $m_2$, a full consistence check requires
at
least going to the $6^{th}$ order (odd orders vanish) in perturbation
- a really
complicated task, as discussed in more details in the appendix. The
second order
does not give any check but fixes a global scale. The fourth order
does contain {\sl some} information: consistency  determines
uniquely
the relation between the masses and $a,b$.
In fact, it is not obvious a priori that this relation will be
physical: finding
it involves solving some quadratic equations whose solutions might
well
be complex, establishing, in fact,  that the TBA is not the right
one. We
have however always found  solutions that are physical, indicating at
least
that the TBA is consistent to that order. Moreover, the general shape
of the
 functions $G$ we obtain can be argued to be the right one based on
limiting
 cases, giving us some confidence in the TBA indeed.

As an example let us consider the first non trivial attractive
reflectionless
case, which corresponds to ${\beta^2 \over 8\pi}={1\over 10}$.
We have numerically solved the TBA equations \rightside, \middi,
\middii\
with $N=2$, $n=2$ for different values of the mass ratio $m /m_1 $.
The coefficients of the analytic expansion
of the running central charge
\eqn\ccexp{
C(r,m/m_1)= 2 + B r^2 + \sum_{k=1}^{\infty} a_{2 k} r^{k (2-g)}}
are determined by fitting.
As a check, the bulk terms of the (extrapolated) limiting cases
$m/m_1 =0$ and
$m_1/m=0$, i.e. the sine-Gordon and the supersymmetric sine-Gordon
points, agree with the exact values  $B=3/\pi$ and $B=0$
within an accuracy of $0.1 \% $.
In Fig.1 we give the result for the adimensional ratio of
coefficients
\eqn\invratio{
I =
12\,\left\{\left[{\Gamma(1/2 +{\beta^2 \over 8\pi})\over \Gamma(1/2
-{\beta^2 \over 8\pi})}
\right]^2 {\Gamma(-{\beta^2 \over 4\pi})\over \Gamma(1+{\beta^2 \over
4\pi})}\right\}^2
{a_4 \over a_2^2}
}
as a function of the mass ratio $m/ m_1$.
\fig{TBA result for the ratio I }{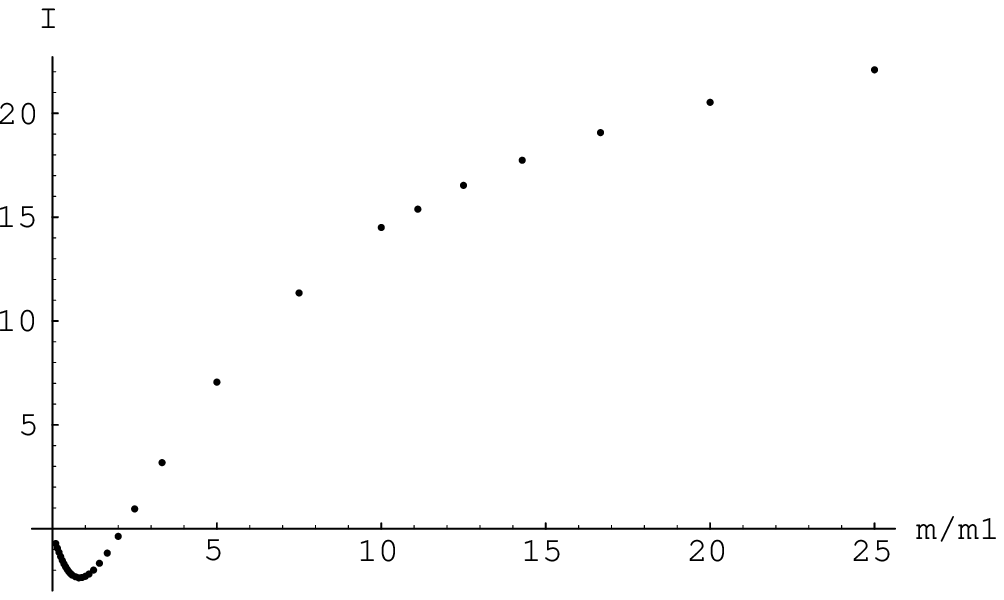}{7cm}
\figlabel\tabb
This curve has to be compared with the same universal ratio
determined in the appendix with
perturbation theory as a function of $x=(k_-/k_+)^2$, shown  in
Fig.2.
\fig{Perturbative result for ratio I}{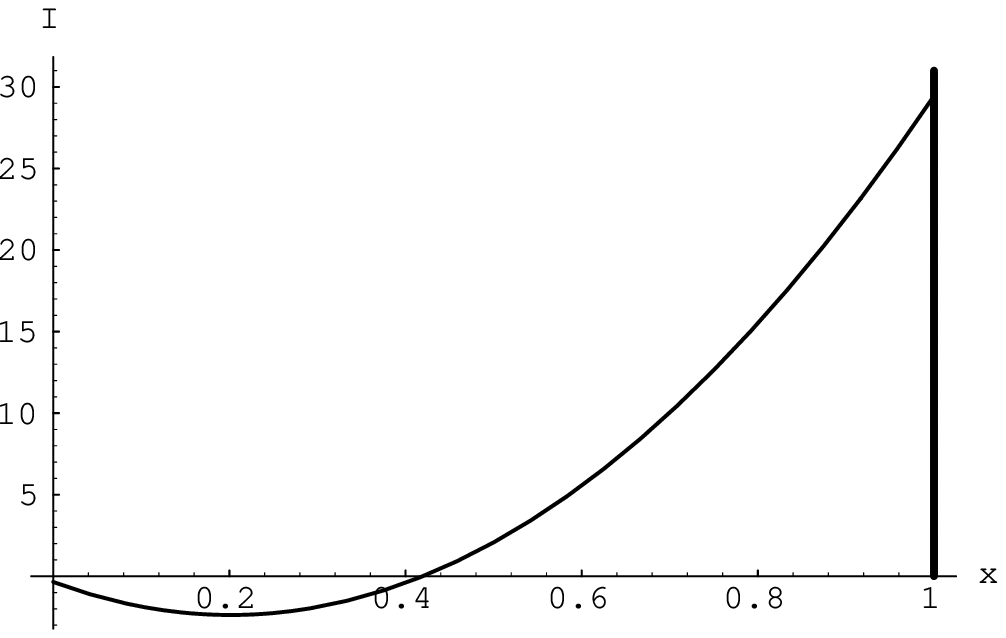}{7cm}
\figlabel\tabb
The limiting  cases are the sine-Gordon model at $x=0$  and the
supersymmetric sine-Gordon  model at $x=1$.
\fig{``Quasi- classical'' behaviour of the ratio of coupling
contants}{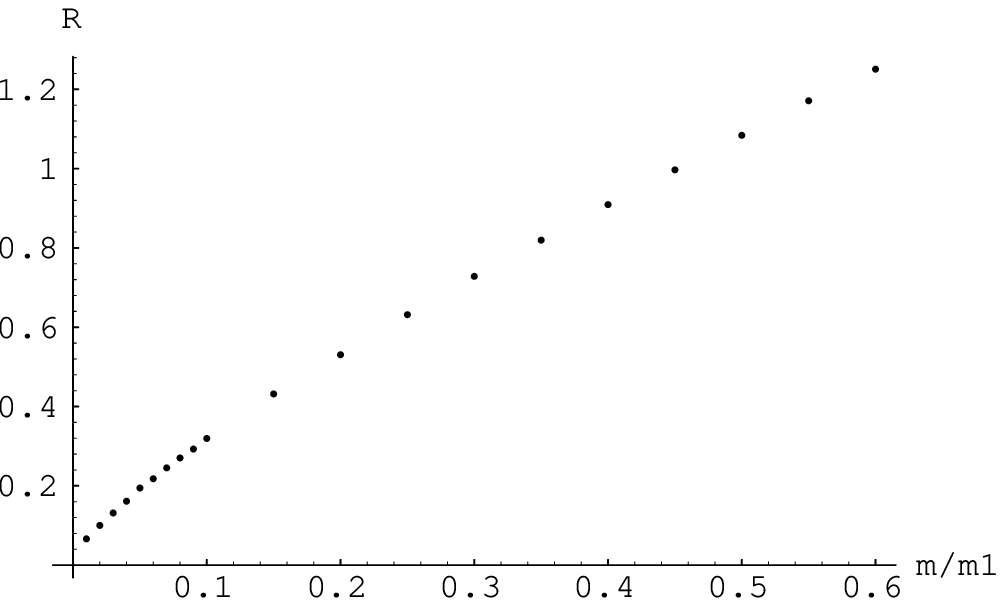}{7cm}
\figlabel\tabb

The value of the universal ratio $I$ at the minimum point found with
perturbation theory is
$I^{pert}_{min}=-2.374$ in good agreement with
the TBA value $I^{TBA}_{min} =-2.366$.
Another important check is the value of the universal ratio at the
supersymmetric sine-Gordon point: the perturbative result is
$I^{pert}_{susy}=29.45$ while the TBA value is
$I^{TBA}_{susy}=29.49$,
confirming therefore the standard analysis\foot{That is, no counter
term
is necessary to make the action supersymmetric and integrable away
from $\beta=0$.}
 of this point

By solving  the second order equation in $(k_-/k_+)^2$ one can
extract the dependence of the coupling constants on the mass ratio.
In Fig.3  we show that the  quantity $R=(b^2 - a^2)/a^2$ is still,
for our value $\beta^2/8\pi=1/10$, almost
a linear function of $m/m_1$, like in the classical limit
 $\beta^2 /8\pi \rightarrow 0$ where $R={m\over m_1}$: moreover, the
slope
is very close to its classical value of $1/2$.

\newsec{Relations with multispin integrable lattice models}

The conjectured TBA appears very naturally in an a priori different
context: the study of inhomogeneous integrable lattice models of XXZ
type
with a mixture of different representations. To explain this in a
concise manner, we refer the reader to \ref\RS{N. Yu Reshetikhin, H.
Saleur, Nucl. Phys. B419 (1994) 507.}, and use similar notations
(though the
matter is quite standard). We
consider thus
an integrable model based on $sl_{q_0}(2)$ R-matrices, whose
``vertical space'' is an array with spins $s_1=s_2=1, s_3=s_4=2,
\ldots, s_{2N-1}=s_{2N}=N$ and $s_{2N+i}=s_i$ otherwise,
ie
made of blocks representing the $N$ first values of $SU(2)$ spin. The
associated spectral parameters alternate
$u_1=-u_2=i{\Lambda+\lambda_1\over 2},
\ldots, u_{2N-1}=-u_{2N}= i{\Lambda+\lambda_N\over 2}$, and
$u_{2N+i}=u_i$
otherwise. The anisotropy is determined by the quantum group
parameter $q_0=\exp\left({i\pi\over \gamma+N}\right)$. For
hamiltonian we chose
\eqn\hamil{\eqalign{H=&{-1\over t}{d\over du} \ln\left\{
t^1\left(i{\Lambda+\lambda_1\over 2}+u\right)
\left[t^1\left(-i{\Lambda+\lambda_1\over
2}-u\right)\right]^{-1}\right.\cr
&\left.\left.\ldots
t^N\left(i{\Lambda+\lambda_N\over 2}+u\right)
\left[t^N\left(-i{\Lambda+\lambda_N\over
2}-u\right)\right]^{-1}\right\}\right|_{u=0}.\cr}}
Here, $t^s$ denotes the transfer matrix based on the foregoing
vertical space and a ``horizontal space'' is a representation of spin
$s$. The whole geometry can be illustrated on the following
picture

\bigskip
\noindent
\centerline{\hbox{\rlap{\raise30pt\hbox{$\hskip.75cm
u_1\ldots\hskip.8cm
u_j\ldots$}}
\rlap{\raise10pt\hbox{$\hskip1.cm\Big|\hskip1.cm\Big|\hskip1.cm\Big|
\hskip1.cm\Big|$}}
\rlap{\lower10pt\hbox{$\hskip.88cm\Big|\hskip1.cm\Big|\hskip1.cm\Big|
\hskip1.cm\Big|$}}
\rlap{\lower30pt\hbox{$\hskip.6cm s_1\ldots\hskip1.3cm s_j\ldots$}}
$\hskip-.4cm u$    -------------------------------------------$
\hskip.5cm s$}}

\bigskip
\noindent

In the case of an array with a single type of spin $j$, the
physical equations are well known to be encoded
in a TBA
 identical to what we studied above for ${\beta^2\over
8\pi}={\gamma\over N(\gamma+N)}$, but  with a mass term on node $j$
only. In this more
general case, it is straightforward to show that one gets now a mass
term
for each of the $N$ first nodes, with masses
\eqn\massform{m_j=M\exp\left(-{(\gamma+N)\over 2}\lambda_j\right),}
where $M={4\over \Delta}\exp\left(-{(\gamma+N)\over
2}\Lambda\right)$.

By taking the continuum limit, this  local integrable lattice model
will give rise to an  integrable quantum field theory that has
exactly   the TBA
conjectured in section 3. This indicates that this TBA  is more than
an
abstract set of equations, but must be related to a genuine quantum
field theory: we   conjecture this field theory is nothing but \act.

It is interesting to observe that the same TBA would also be obtained
by chosing
a uniform spectral parameter (eg all $\lambda_j=0$) but by putting
different
amounts of the various spins, with densities proportional to the
masses $m_j$;
see \ref\VMN{H.J. de Vega, L. Mezincescu and R. I. Nepomechie,
J.Mod.Phys. B8 (1994) 3473-3485.} for more details on this approach.
In the latter
reference in particular, S matrices are directly derived from the
lattice
regularization, and agree with the results of section 3. In the
context of lattice models, central charges have also been computed
with TBA similar to ours \ref\VMNI{H.J. de Vega, L. Mezincescu and
R.I. Nepomechie, Phys. Rev. B49
(1994) 13223.},\ref\AM{S.R. Aladin and M.J. Martins,
J.Phys.A:Math.Gen. 26 (1993) L529.}

\newsec{Impurity problems}

The same argument of perturbative integrability carries through in
the
case of impurity problems \ref\GZ{S. Goshal, A. B. Zamolodchikov
Int.J.Mod.Phys.A9 (1994) 3841, Erratum-ibid. 4353.}. Consider
thus the problem with free bosons $\Phi_i,\Phi$ in the bulk, and
an  interaction term at the boundary
\eqn\bdr{H_{bdr}=\left(\sum_{j=1}^{N} a_j\Psi^{(j)}(0)\right)
 S^- e^{i\beta\Phi
(0)}+\hbox{ conjugate}.}
Here, $S^\pm$ are raising and lowering operators in a spin
$j$\foot{In our conventions,
the fundamental has spin one.}
representation
of the quantum group $sl(2)_{q_0}$ (here the deformation parameter
is not the same than the one appearing in the S-matrices earlier, but
rather
the one of the lattice model above,
 $q_0=\exp\left( {i\pi\over \gamma+N}\right)$).  Note that only the
right moving part of the fields appears in the action, but that the
bosons $\phi,\Phi$ all have
Neumann boundary conditions, ie their right and left moving
components
are identical at the boundary. Alternatively, one could thus express
the boundary perturbation with the total fields, and exponentials of
half the argument. A particular case is where the boundary spin is in
a ``cyclic''
representation, and can be gauged away
\ref\FS{P. Fendley, H. Saleur, Phys. Rev. Lett. 75 (1992) 4495.},
\ref\BLZ{V. Bazhanov,
A. Lukyanov, A. B. Zamolodchikov, Commun.Math.Phys.177 (1996).}. One
finds then
\eqn\bdr{H_{bdr}=\sum_{j=1}^{N}
\lambda_j\exp i\left(\phi_j+\beta\Phi\right)+\hbox{ conjugate},}
with $\lambda_j$ defined in \lamdef.

In the simplest case of the repulsive regime, and for $\gamma$ an
integer,
the boundary free energy for spn $j$ with
$j\leq \gamma+N-2$ reads, as in the usual
anisotropic Kondo problem ,
\eqn\fdrb{F_{bdr}=-T\int {d\theta\over 2\pi}{1\over
\cosh(\theta-\theta_B)}
\ln\left(1+e^{-\epsilon_{j}}\right).}
Here, the $\epsilon_j$ are obtained by solving the TBA system  \tba\
in the
massless limit; this means, one sends the masses to zero and the
rapidities to
$\infty$, such that only right moving  particles with
dispersion relation $e=p$ remain. In the TBA, the source terms are
obtained by the simple substitution  $\cosh\theta\to {1\over
2}e^{\theta}$.
The ``masses'' are simply parameters with the physical dimension of
an inverse length. The rapidity $\theta_B$ is such that
$m_1e^{\theta_B}\propto a_1^{ N+\gamma\over 2}$. For the case
of the cyclic spin, the free energy reads as \fdrb\ but with
$j=\gamma+N$.

It is especially interesting to consider the free energy in the
classical  limit \foot{This is usually called the ``Toulouse'' limit
in the
context of impurity problems \ref\Tou{G. Toulouse, C.R. Acad. Sci.
268 (1969) 1200.}}
. There, the same formula \fdrb\ holds for
spin $j\leq N-2$ and the TBA now given by \class.
For the spin $j=N-1$, we have, due to some of the foregoing changes
of variables,
\eqn\gati{
F_{bdr}=-T\int {d\theta\over 2\pi}{1\over \cosh(\theta-\theta_B)}
\left[\ln\left(1+e^{-\epsilon_{N-1}}\right)
+\ln\left(1+e^{-\epsilon_N}\right)\right].}
Finally, for the case of cyclic boundary spin.
\eqn\gatii{
F_{bdr}=-T\int {d\theta\over 2\pi}{1\over \cosh(\theta-\theta_B)}
\ln\left(1+e^{-\epsilon_N}\right),}
(in the last formulas,  we have subtracted the trivial boundary free
energy
of the free boson $\Phi$).

In the case $N=2$, the problem has been studied in
\ref\FGN{M.Fabrizio, A. Gogolin, P. Nozieres, cond-mat/9412118}. Take
an impurity of spin $1/2$, and
set $S^+=d^+, S^-=d$. Introducing
the Dirac fermion $\Upsilon=\psi+i\chi$, the boundary action reads
$$
{a-b\over 2}\left(\Upsilon^+d^++\Upsilon d\right)+{a+b\over 2}
\left(\Upsilon d^++\Upsilon^+ d\right)
$$
According to \gati, the free energy in the spin
$1/2$ case (ie $j=1$ in our notations), reads, using notations of
section 2,
$$
F_{bdr}=-T\int {d\theta\over 2\pi}\left[{1\over
\cosh(\theta-\theta_B)}
+{1\over \cosh(\theta-\theta'_B)}\right]
\ln\left(1+e^{-e^{\theta}/T}\right),
$$
where we have made a shift of the variable of integration, and
thus ${e^{\theta_B}/e^{\theta_B'}}\propto {a^2\over b^2}$.
In terms of the $\lambda,\mu$ variables describing the couplings to
the
different
channels, this reads ${e^{\theta_B}/e^{\theta_B'}}\propto
\left({\lambda-\mu\over \lambda+\mu}\right)^2$, in
agreement
with results of \FGN.

For general $N$, and when  all the masses $m_k$ but
$m_N$ vanish (
the standard generalized supersymmetric case) we obtain from \gatii\
the ratio of degeneracy factors in the UV and IR ${g_{UV}\over
g_{IR}}=\sqrt{N}$, corresponding to a flow from free to fixed
boundary conditions in the $Z_N$ model. In general,  we have here
the solution of a multiboson problem with arbitrary couplings; when
$N=3$ for
instance, this means that the problem with boundary perturbation
$$
\lambda\cos\phi_1+\mu\left(\cos\phi_2+\cos\phi_3\right)
$$
is integrable. Rexpressing the bosons $\phi$ in terms of the
independent
bosons $\Phi$, we obtain a two boson problem that is nothing
but a quantum wire problem: see \AOS\ for more details.

\newsec{Conclusions}

We feel there is more to understand in the theories we have
addressed.
The relation with the bare Bethe
ansatz solutions of \NatanI, \NatanII\ in the color isotropic case
is poorly understood, as discussed in the text. The scattering theory
we have
proposed is rather mysterious,
and we have not answered the question of what the scalar particles
have to do
with the flavor symmetry breaking. The problem of analytically
determining
the relation between the action parameters $a_j$ and the masses $m_j$
remains  in
general open. Finally, we have restricted to positive coefficients
$a_j$ in
the problem, but it is  clear that the perturbations will not
always be massive if we allow some of these coefficients to be
negative: where the
theories flow to in that case is also an open question.

\vskip4pt
\noindent{\bf Acknowledgments}: We thank N. Andrei F. Lesage, and
Al. Zamolodchikov
for many useful discussions.

\newsec{Appendix}

As a check of the correctness of our solution we now compute with
conformal perturbation theory the free energy for the case $N=2$.
The result has to be compared  with the free energy obtained by
numerically solving
the TBA equations.
Let's consider the system in the strip geometry $(R,L)$ defined by
the action
$A = A_{cft} - \int_{strip}\Phi_{int} $, where the interaction is
given by \act.
The dimensionless running central charge
\eqn\ccdef{
C(R,a,b) = \lim_{L\rightarrow \infty} {6 R\over\pi L} \ln Z[R,L] =
{6 R^2\over\pi} F
}
becomes in perturbation theory
\eqn\ccpert{
C(R,a,b) = c_{UV} + 12 \lim_{L\rightarrow \infty} {R\over 2\pi L}
\sum_{k=2}^{\infty} {1\over k!}\int_{strip}d^2 w_1 \cdots d^2  w_k
<\Phi_{int}(w_1) \cdots\Phi_{int}(w_k) >^c .
}
The first correction $k=2$ is  ultraviolet divergent for any real
value
of $\beta^2$, the anomalous dimension of the perturbing operator
being $\Delta =1/2 +{\beta^2 \over 8\pi}$. Whatever regularization we
choose, for example a
radially ordered one \ref\LC{A.W.W. Ludwig and J.L. Cardy, Nucl.Phys.
B285[FS19] (1987) 687.}, we obtain a divergent part, to be subtracted
by a constant counterterm in the lagrangian, and a universal part.
The counterterm contains a possible finite contribution giving rise
to
a non-universal bulk term which has to be fixed by a normalization
condition.
This condition is $C(R,a,b)\rightarrow 0$ when $R\rightarrow \infty$.
In practice the bulk term will be determined by comparison with the
TBA result, while in integrable theories with only one mass scale it
can be computed
analytically.
Taking into account the first non trivial correction, the running
central charge reads
\eqn\cctwol{\eqalign{
C(R,a,b)= & c_{UV} + c_{bulk}(R,a,b) \,\,\,\,  +  \cr
& 6 (2\pi )^{\beta^2 \over 2\pi} \left[{\Gamma(1/2 +{\beta^2 \over
8\pi})\over
\Gamma(1/2 -{\beta^2 \over 8\pi})}\right]^2
{\Gamma(-{\beta^2 \over 4\pi})\over \Gamma(1+{\beta^2 \over 4\pi})}
\left[(a^2 + b^2) R^{1-{\beta^2 \over 4\pi}}\right]^2 .\cr
}}
The theory depends  on the pair of massive coupling constants
$(a,b)$, or alternatively on the two masses $(m1,m)$. Since we are
going to compare
our perturbative expansion with the TBA result it is useful to
introduce the dimensionless coupling constants $(k_{+},k_{-})$ and to
make explicit
the dependence on the mass ratio:
$a^2+b^2 = k_+(m/m_1) \, m^{1-{\beta^2 \over 4\pi}}$  and $a^2 -b^2 =
k_-(m/m_1) \,m^{1-{\beta^2 \over 4\pi}}$.
Defining the dimensionless quantity $r=m R$, the second order
running central charge becomes
\eqn\cctwom{\eqalign{
C(r,m/m_1 )=& 2 + c_{bulk} (m/m_1 ) \,r^2 \,\,\,\,+ \cr
& 6 \,(2\pi )^{\beta^2 \over 2\pi} \left[{\Gamma(1/2 +{\beta^2 \over
8\pi})\over \Gamma(1/2 -{\beta^2 \over 8\pi})}\right]^2
{\Gamma(-{\beta^2 \over 4\pi})\over \Gamma(1+{\beta^2 \over 4\pi})}
\left[k_+(m/m_1 )\right]^2 \,r^{2-{\beta^2 \over 2\pi}} .\cr
}}

Let's consider  the case $\beta^2 < 2\pi$, which includes the
attractive regime. The only UV divergence of perturbation theory is
the one that occurs at second order. All the other perturbative
contributions are  UV finite.
The third order correction is zero because the unperturbed three
points
correlation function of the interaction is zero by charge neutrality.
In order to compute the fourth order correction we map the $w$-strip
onto the
$z$-plane, $z=\exp(i 2\pi w/R)$, and  we express by Wick theorem the
unperturbed four points correlation function
\eqn\unpcf{
G=2\left|{(a^2-b^2)^2 \over z_{12} z_{34} } -{(a^2+b^2)^2 \over
z_{13} z_{24} }
+{(a^2+b^2)^2 \over z_{14} z_{23} }\right|^2 \,
\left[{\left|z_{12}\right|\left|z_{34}\right|\over
\left|z_{13}\right|\left|z_{14}\right|\left|z_{23}\right|\left|z_{24}\right|}
\right]^{\beta^2 \over 2\pi} +\hbox{Perm.}
}
to which we have to subtract the disconnected term
\eqn\disc{
D=4 (a^2+b^2)^4 \left\{
\left[{1\over
\left|z_{12}\right|\left|z_{34}\right|}\right]^{2+{\beta^2 \over
2\pi}}
+\left[{1\over
\left|z_{13}\right|\left|z_{24}\right|}\right]^{2+{\beta^2 \over
2\pi}}
+\left[{1\over
\left|z_{14}\right|\left|z_{23}\right|}\right]^{2+{\beta^2 \over
2\pi}}\right\}.
}

It is useful to rewrite the correlation function as the sum of
two pieces
\eqn\sumonecf{
G_1 =2\left[{(a^2-b^2)^4 \over |z_{12}|^2 |z_{34}|^2 } +{(a^2+b^2)^4
\over |z_{13}|^2 |z_{24}|^2 }
+{(a^2+b^2)^4 \over |z_{14}|^2 |z_{23}|^2 }\right] \,
\left[{\left|z_{12}\right|\left|z_{34}\right|\over
\left|z_{13}\right|\left|z_{14}\right|\left|z_{23}\right|\left|z_{24}\right|}
\right]^{\beta^2 \over 2\pi} +\hbox{Perm.}
}
and
\eqn\sumtwocf{\eqalign{
G_2=& 2\left[-{(a^4-b^4)^2 \over z_{12} z_{34} z_{13}^* z_{24}^* }
+{(a^4-b^4)^2 \over z_{12} z_{34} z_{14}^* z_{23}^*}
-{(a^2+b^2)^4 \over z_{13} z_{24} z_{14}^* z_{23}^*
}+\hbox{c.c.}\right] \,
\left[{\left|z_{12}\right|\left|z_{34}\right|\over
\left|z_{13}\right|\left|z_{14}\right|\left|z_{23}\right|\left|z_{24}\right|}
\right]^{\beta^2 \over 2\pi} \cr
& \,\,\,+\,\,\,\hbox{Perm.}
}}
and to compute the perturbation integral as the sum of the two
corresponding
integrals $\int(G_1-D)$ and $\int G_2$, both separately UV-finite.
As a result, the fourth order term is the sum $C^{(4)}=C_1^{(4)}+
C_2^{(4)}$
where
\eqn\conedef{
C_1^{(4)}=12\lim_{L\rightarrow\infty}{R\over  2\pi L}\int_{strip}
d^2w_1\cdots d^2w_4 \left.\left[G_1 -D\right]\right|_{strip}
}
and
\eqn\ctwodef{
C_2^{(4)}=12\lim_{L\rightarrow\infty}{R\over  2\pi L}\int_{strip}
d^2w_1\cdots d^2w_4 \left.G_2\right|_{strip}\,.
}

In the computation of the first integral we take the limit
$L\rightarrow\infty$
first, and we cancel the overall volume of the strip $R L$, and then
we map
the infinite strip to the whole $z$-plane. We thus get
\eqn\coneint{
C_1^{(4)}={(2\pi )^{\beta^2 \over \pi} \over 2} R^{4-{\beta^2 \over
\pi}} \int {d^2 z_2\over 2\pi}
{d^2 z_3\over 2\pi} {d^2 z_4\over 2\pi} \left.\left(|z_2| |z_3| |z_4|
\right)^{{\beta^2 \over 4\pi}-1} \left[G_1 - D\right]\right|_{z_1 =
1}
}
and by using the residual symmetry on the integration variables we
are essentially reduced to two integrals
\eqn\conetot{
C_1^{(4)}=  3 (2\pi )^{\beta^2 \over \pi} R^{4-{\beta^2 \over \pi}}
\left[ (a^2 - b^2)^4 A_1 +  (a^2 +b^2)^4 A_2\right]
}
where
\eqn\aoneint{
A_1=\int {d^2 z_2\over 2\pi}
{d^2 z_3\over 2\pi} {d^2 z_4\over 2\pi} \left.{\left(|z_2| |z_3|
|z_4|
\right)^{{\beta^2 \over 4\pi} -1} \over |z_{12}|^2 |z_{34}|^2}
\left[{\left|z_{12}\right|\left|z_{34}\right|\over
\left|z_{13}\right|\left|z_{14}\right|\left|z_{23}\right|\left|z_{24}\right|}
\right]^{\beta^2 \over 2\pi} \right|_{z_1 = 1}
}
and
\eqn\atwoint{\eqalign{
A_2=2\int {d^2 z_2\over 2\pi}
{d^2 z_3\over 2\pi} {d^2 z_4\over 2\pi} & {\left(|z_2| |z_3| |z_4|
\right)^{{\beta^2 \over 4\pi} -1} \over |z_{12}|^2 |z_{34}|^2} \times
\cr
&\left.
\,\,\,\,\left\{\left[{\left|z_{13}\right|\left|z_{24}\right|\over
\left|z_{12}\right|\left|z_{14}\right|\left|z_{23}\right|\left|z_{34}\right|}
\right]^{\beta^2 \over 2\pi} -\left[{1\over |z_{12}|
|z_{34}|}\right]^{\beta^2 \over 2\pi}\right\}\right|_{z_1 = 1}.
}}
These two integrals are of Dotsenko-Fateev type \ref\DF{Vl.S.
Dotsenko and
V.A. Fateev, Nucl.Phys. B240[FS12] (1984) 312.}.
By deforming the integration contours they can be transformed into
products
of two factors. Each factor is the sum, with proper trigonometric
coefficients, of one-dimensional integrals  of this kind
\eqn\intdot{
\int_0^1\prod_{i=1}^3\left[dv_i v_i^{\alpha_i}
(1-v_i)^{\beta_i}\right]
(1-v_1 v_2)^{\gamma_1} (1-v_2 v_3)^{\gamma_2} (1-v_1 v_2
v_3)^{\gamma_3}.
}
They can be formally integrated  by binomially expanding  the last
three factors of the integrand and using the fundamental integral
$\int_0^1 dv v^a (1-v)^b = \Gamma(1+a)\Gamma(1+b)/\Gamma(2+a+b)$.
Therefore the two integrals $A1$ and $A2$ can be reduced to the
computation
of (products of) converging series of three indices.
We don't give the explicit expressions because of their algebraic
heaviness.
The method is really a straightforward generalization of the paper
\ref\Dot{Vl.S. Dotsenko, Nucl.Phys. B314(1989) 687.}.
The result can now be evaluated numerically by extrapolating the
finite sums.

The integral contributing to $C_2^{(4)}$ is evaluated with the method
developed
in \ref\SI{H. Saleur and C. Itzykson, J.Stat.Phys. 48 (1987) 449.}.
We first map the finite strip to the annulus $\rho < |z| < 1$, where
$\rho=\exp(-2\pi L/R)$, and then we compute the leading contribution
of the integral in the limit $\rho\rightarrow 0$. Using the symmetry
under
permutation of  the four integration variables we obtain
\eqn\ctwoint{
C_2^{(4)}=12 (2\pi )^{\beta^2 \over \pi} R^{4-{\beta^2 \over \pi}}
\lim_{\rho\rightarrow 0} {1\over\ln({1\over\rho})}
\int_{\rho}^1dr_4\int_{\rho}^{r_4}dr_3 \int_{\rho}^{r_3}dr_2
\int_{\rho}^{r_2}dr_1 \prod_{i=1}^4\int_0^{2\pi}{d\theta_i\over 2\pi}
r_i^{\beta^2 \over 4\pi}
\,G_2\,.
}
Having ordered the integration variables it is now possible to
binomially expand each factor $z_{jl}^{\gamma}=r_j ^\gamma
\exp(i\theta_j\gamma) (1 -\exp(i(\theta_l-\theta_j)
r_l/r_j)^{\gamma}$ for $j>l$.
Then we obtain a series on $12$ indices constrained by three
independent conditions from the angular integrations, each term of
the series
being a product of binomial coefficients and four radially ordered
integrals of powers of the $r_i$. The last integral, the one in
$dr_4$, gives the necessary overall volume divergence $\prop
\ln(1/\rho)$.
The result is
\eqn\ctwotot{
C_2^{(4)}= 3 (2\pi )^{\beta^2 \over \pi} R^{4-{\beta^2 \over \pi}}
\left[ (a^4 - b^4)^2 S_1 +  (a^2 +b^2)^4 S_2\right]
}
where we give as an example a term contributing to $S_2$
\eqn\example{
{\sum_{n_1\ldots ,m_1\ldots}}'  {- 2 q_{n,m}\over
 (n_1 +n_2+n_3+{1\over 2} +{\beta^2 \over 8\pi})
(n_3 +n_5+n_6+{1\over 2} +{\beta^2 \over 8\pi}) (n_2+n_3+n_4+n_5+ 1
+{\beta^2 \over 2\pi}) }
}
where the symbol $\sum'$ means the following conditions on the
indices
$n_1,\ldots n_6, m_1, \ldots m_6 $
\eqn\constrexa{
\eqalign{n_1 +n_2 +n_3 &= m_1 +m_2 +m_3  \,\,\,\hbox{;}\,\,\,\,\,\,\,
n_1 -n_4 -n_5 = m_1 -m_4 -m_5 \cr
n_2 +n_4 -n_6 &=m_2 +m_4 -m_6 \,\,\,\hbox{;}\,\,\,\,\,\,\,
n_3 +n_5 +n_6 = m_3 +m_5 +m_6 .\cr}
}
The coefficient $q_{n,m}$ is
\eqn\exampq{
\eqalign{
q_{n.m}=&
b_{n_1}\left(1+{\beta^2 \over 4\pi}\right) b_{m_1}\left(1+{\beta^2
\over 4\pi}\right) b_{n_2}\left(-{\beta^2 \over 4\pi}\right)
b_{m_2}\left(1-{\beta^2 \over 4\pi}\right) b_{n_3}\left(1-{\beta^2
\over 4\pi}\right) b_{m_3}\left(-{\beta^2 \over 4\pi}\right)
\cr &
b_{n_4}\left(1-{\beta^2 \over 4\pi}\right) b_{m_4}\left(-{\beta^2
\over 4\pi}\right) b_{n_5}\left(-{\beta^2 \over 4\pi}\right)
b_{m_5}\left(1-{\beta^2 \over 4\pi}\right) b_{n_6}\left(1+{\beta^2
\over 4\pi}\right) b_{m_6}\left(1+{\beta^2 \over 4\pi}\right) \cr }
}
with $b_n(x)= \Gamma(n+1-x)/\left(n! \Gamma(1-x)\right)$.

Summarizing, the fourth order correction for the running central
charge
is given by
\eqn\ccfourth{
C^{(4)}=3 (2\pi )^{\beta^2 \over \pi} \left[\left(A_2 +S_2\right)
k_+^4 + S_1 k_-^2 k_+^2
+ A_1 k_-^4 \right] r^{4 - {\beta^2 \over \pi}}
}
where the numbers $A_1$,$A_2$,$S_1$,$S_2$ depend on ${\beta^2 \over
8\pi}$ and can be computed
numerically by extrapolating the values of the respective finite
sums.
Unfortunately the above sums, especially the $A_i$ ones, are slow to
converge affecting therefore the precision of the extrapolated
values.
For the most favourable case ${\beta^2 \over 8\pi}={1\over 10}$
the extrapolated values are
\eqn\extrap{\eqalign{
A_1=&49.9 \,\,\,\,,A_2=1.44 \cr
S_1=&-20.1\,\,\,\,,S_2=-1.79\,\,\,\,.}
}
For the first two coefficients $A_1$ and $A_2$ we have used the
VBS extrapolation method over the set of finite sums with $25< N\leq
40$, while
the other coefficients $S_1$ and $S_2$ have been determined
with the BST extrapolation method with convergence parameter
$\omega =1$ over finite sums with $N\leq 7$.
The need for difference extrapolation methods is due to the
difference
in the rate of convergence of the series.
With our choice the given extrapolated values are the most stable
with
respect to $N$.
A coincise introduction to the above extrapolation methods can be
found in \ref\CH{P.Christe, M.Henkel, ``Introduction to conformal
invariance and its applications to critical phenomena'', Lecture
Notes
in Physics, volume  16, Springer-Verlag.}.

The numerical integration of the TBA equations gives us the running
central charge $C(r,m/m_1)$ with high precision and therefore the
coefficients of the
expansion
\eqn\ccexpapp{
C(r,m/m_1)= 2 + B r^2 + \sum_{k=1}^{\infty} a_{2 k} r^{k (2-{\beta^2
\over 2\pi})}
}
can be determined with a standard fitting procedure.
By matching the perturbative computation with the first two
coefficients $a_2$ and $a_4$ we can determine
now the two functions $k_+(m/m_1)$, $k_-(m/m_1)$
as solution of the  following second order algebraic system
\eqn\matchsystem{
\eqalign{
a_2 &=  6\,(2\pi )^{\beta^2 \over 2\pi} \left[{\Gamma(1/2 +{\beta^2
\over 8\pi})\over \Gamma(1/2 -{\beta^2 \over 8\pi})}\right]^2
{\Gamma(-{\beta^2 \over 4\pi})\over \Gamma(1+ {\beta^2 \over 4\pi})}
k_+^2 \cr
{a_4 \over a_2^2} &= {1\over 12}\,\left\{\left[{\Gamma(1/2 -{\beta^2
\over 8\pi})\over \Gamma(1/2 +{\beta^2 \over 8\pi})}\right]^2
{\Gamma(1+{\beta^2 \over 4\pi})\over \Gamma(-{\beta^2 \over
4\pi})}\right\}^2
\left\{A_1 \left({k_- \over k_+}\right)^4 + S_1 \left({k_- \over
k_+}\right)^2 + (A_2 + S_2) \right\}.}
}

\listrefs

\bye